\documentclass[conference]{IEEEtran}
\IEEEoverridecommandlockouts
\usepackage{cite}
\usepackage{amsmath,amssymb,amsfonts}
\usepackage{algorithmic}
\usepackage{graphicx}
\usepackage{textcomp}
\usepackage{xcolor}

\usepackage[T1]{fontenc}
\usepackage{multirow}
\usepackage[hidelinks]{hyperref}
\usepackage{caption}
\usepackage{subcaption}
\usepackage{textgreek}

\def\BibTeX{{\rm B\kern-.05em{\sc i\kern-.025em b}\kern-.08em
    T\kern-.1667em\lower.7ex\hbox{E}\kern-.125emX}}
\begin{document}

\title{Efficient RDF Streaming for the Edge-Cloud~Continuum
\thanks{This work is part of the ASSIST-IoT project that has received funding from the EU’s Horizon 2020 research and innovation programme under grant agreement No 957258.}
}

\author{\IEEEauthorblockN{Piotr Sowiński}
\IEEEauthorblockA{\textit{Systems Research Institute} \\
\textit{Polish Academy of Sciences}\\
Warsaw, Poland \\
0000-0002-2543-9461}
\and
\IEEEauthorblockN{Katarzyna Wasielewska-Michniewska}
\IEEEauthorblockA{\textit{Systems Research Institute} \\
\textit{Polish Academy of Sciences}\\
Warsaw, Poland \\
0000-0002-3763-2373}
\and
\IEEEauthorblockN{Maria Ganzha}
\IEEEauthorblockA{\textit{Systems Research Institute} \\
\textit{Polish Academy of Sciences}\\
Warsaw, Poland \\
0000-0001-7714-4844}
\and
\IEEEauthorblockN{Wiesław Pawłowski}
\IEEEauthorblockA{\textit{Dept. of Mathematics, Physics, and Informatics} \\
\textit{University of Gdańsk}\\
Gdańsk, Poland \\
0000-0002-5105-8873}
\and
\IEEEauthorblockN{Paweł Szmeja}
\IEEEauthorblockA{\textit{Systems Research Institute} \\
\textit{Polish Academy of Sciences}\\
Warsaw, Poland \\
0000-0003-0869-3836}
\and
\IEEEauthorblockN{Marcin Paprzycki}
\IEEEauthorblockA{\textit{Systems Research Institute} \\
\textit{Polish Academy of Sciences}\\
Warsaw, Poland \\
0000-0002-8069-2152}
}

\maketitle

\begin{abstract}
With the ongoing, gradual shift of large-scale distributed systems towards the edge-cloud continuum, the need arises for software solutions that are universal, scalable, practical, and grounded in well-established technologies. Simultaneously, semantic technologies, especially in the streaming context, are becoming increasingly important for enabling interoperability in edge-cloud systems. However, in recent years, the field of semantic data streaming has been stagnant, and there are no available solutions that would fit those requirements. To fill this gap, in this contribution, a novel end-to-end RDF streaming approach is proposed (named \textsc{Jelly}). The method is simple to implement, yet very elastic, and designed to fit a wide variety of use cases. Its practical performance is evaluated in a series of experiments, including end-to-end throughput and latency measurements. It is shown that \textsc{Jelly} achieves vastly superior performance to the currently available approaches. The presented method makes significant progress towards enabling high-performance semantic data processing in a wide variety of applications, including future edge-cloud systems. Moreover, this study opens up the possibility of applying and evaluating the method in real-life scenarios, which will be the focus of further research.

\end{abstract}


\section{Introduction}
\label{sec:intro}

Nowadays, a gradual shift in IoT ecosystems can be observed, with the ``traditional'' division of computing infrastructures into cloud, edge, far edge, etc., being replaced with the emerging idea of \emph{edge-cloud continuum}~\cite{eCloud}. It provides a new way of thinking about how heterogeneous devices are to cooperate seamlessly within large-scale ecosystems. However, for this concept to become reality, a paradigm shift is needed in how software and interfaces are designed~\cite{seamless-edge-cloud}. For example, when one focuses on the solution being performant on microcontrollers, usually the method's scalability and compatibility with existing interfaces is sacrificed. This kind of approach is no longer appropriate, as elastic and adaptable solutions are needed, which scale vertically and horizontally, and can be implemented effortlessly on a wide variety of hardware and software platforms.

Semantic technologies have proven themselves to be a vital asset for enabling interoperability in IoT, and facilitating data reuse~\cite{ganzha2017semantic,rahman2020comprehensive}. However, the practicality of semantics-based solutions is often criticized, especially with regard to their implementation complexity and performance~\cite{ganzha2017towards}. Both these issues can, quite obviously, hamper the application of semantics in edge-cloud systems. In this work, the focus is on a specific aspect of semantic technologies, namely the streaming of semantic data. There are many ways in which \textit{semantic data streaming} can be defined, however, the following definition was chosen, due to it being very universal and able to cover most use cases. Semantic data streaming is a one-way communication act between two actors: the producer and the consumer. The producer sends a potentially unbounded sequence of Resource Description Framework (RDF)~\cite{Cyganiak:14:RCA} statements (triples or quads) to the consumer over the network. The protocol used for network communication varies from application to application and can span from synchronous, point-to-point approaches (such as gRPC), to asynchronous, distributed methods, involving a message broker (such as Apache Kafka). For practical reasons, a stream of RDF statements is divided into discrete RDF graphs (messages, blocks) to be sent over the network. These messages are serialized by the producer and then deserialized by the consumer. To improve communication efficiency, the messages can be compressed, either by the serialization format itself, or at the network protocol level (using, for example, gzip compression).

Stream processing of semantic data is a key problem in implementing semantics-enabled distributed systems, being used in a wide variety of applications~\cite{ganzha2017towards, tommasini2021web, kabisch2015standardized}. However, upon a closer examination of the available solutions for semantic data streaming (see Section~\ref{sec:bg}), it becomes clear that they are hardly adequate to meet the requirements of future edge-cloud continuum systems. For instance, up until now, there has been no established standard for RDF data streaming. Moreover, the existing solutions are fragmented, of little practical use, and often focused on a very specific application domain.

This is the research gap that this contribution seeks to cover. The proposed solution aims to be all-encompassing with regard to the outlined requirements, by being scalable, performant, easy to implement, and based on already established best practices, used widely by both IoT and cloud industries. The implementation of the approach, results, and other additional materials are available publicly under an open-source license\footnote{\url{https://github.com/Ostrzyciel/jelly-wfiot-2022}}~\cite{materials}.

\section{Related Works}\label{sec:bg}

In the literature, one can encounter several approaches to defining what an RDF stream is. The definition given in Section~\ref{sec:intro} does not apply directly to all of the works presented below. For instance, some of them are only appropriate for streams of RDF graphs, not statements. Also worth mentioning is the definition used by the RDF Stream Processing community, which focuses on \emph{timestamped} RDF streams, where data is annotated with the time, at which the event occurred and/or was processed~\cite{tommasini2021rsp4j}. However, the definition used in this work can be argued to be more general, as timestamped streaming can be achieved by the inclusion of additional RDF statements with the required timestamps.

One of the earlier works tackling efficient RDF streaming is the introduction of the Streaming HDT (S-HDT) format~\cite{hasemann2012rdf} -- a lightweight binary RDF representation that is oriented towards resource-constrained IoT devices. The format features a simple, dictionary-based compression method. Additionally, a custom node-to-node communication protocol is introduced. Unfortunately, the authors did not publish their source code. Moreover, no additional resources or mentions of \mbox{S-HDT's} use in practice could be found. \mbox{S-HDT} builds on the earlier work on HDT~\cite{fernandez2013binary}, which uses a more advanced compression method, but can only work with entire datasets. Hence, it is unsuitable for streaming applications.

RDSZ~\cite{fernandez2014rdsz} was the first RDF streaming approach that attempted to significantly lower the size of the serialized representation. The method uses the standard Turtle format as its base, but it also can apply differential encoding in cases that can benefit from it. The output is then additionally compressed with Zlib (the Deflate algorithm). However, the method seems rather impractical, being over two times slower than the Turtle baseline, in serialization and deserialization.

A more elaborate solution to the same problem is the Efficient RDF Interchange (ERI) format~\cite{fernandez2014efficient}. The algorithm focuses on achieving competitive compression ratios through the use of dictionaries, storing structural information about the graph. ERI assumes that the stream is divided into ``blocks'' of triples (usually 4096 in length), with each block being processed atomically. However, this is also an important limitation of this approach, as it cannot effectively handle small blocks, or individual triples/quads. Moreover, the implementation is not integrated with any RDF library (such as Apache Jena~\cite{carroll2004jena}).

Another notable approach is the RDF serialization based on the binary W3C Efficient XML Interchange (EXI) format~\cite{kabisch2015standardized}. It targets Web of Things environments and industrial devices that, typically, need to exchange measurement data. RDF EXI has the ability to integrate a base ontology (TBox only) into the protocol with the use of code generation, which can greatly reduce the size of the messages. This is done at the cost of interoperability. Specifically, both the producer and the consumer need to be aware of the ontology that is being used, and have the necessary code already generated. The authors of the format claim that one of its biggest advantages is that it is based on a W3C standard (EXI). However, the implementation for the RDF serialization was not open-sourced and (possibly for this reason) the format has seen little adoption.

Over the years, the RDF Stream Processing community has largely focused on streaming in the context of query languages, query execution engines, and reasoners~\cite{tommasini2021rsp4j}, for timestamped RDF streams. However, few approaches to RDF data streaming were proposed. Worth mentioning is CQUELS-Cloud, a distributed RDF stream processing framework~\cite{le2013elastic} that internally compresses the RDF data, using dictionary encoding. However, also in this case, the code was not published, and the protocol is only used internally, within the stream processing cluster.

Overall, in the past, research on RDF streaming has been very limited. Several works~\cite{fernandez2014efficient,fernandez2014rdsz,kabisch2015standardized} focused on achieving high compression ratios, with little regard for the practical applicability of the proposed solutions. The throughput measurements were only performed on parts of the pipelines (such as serialization and/or deserialization) in isolation, which is a poor indicator of real-life performance. It should be also noted that the mentioned works have been published in 2012--15 with no follow-up that could be found.

Besides the mentioned works, it is also worth pointing out that two non-standard binary serializations were implemented in Apache Jena~\cite{jena-binary} -- using Apache Thrift (2014) and, more recently, Google Protocol Buffers (2021). These are very simple serialization formats, with no compression built-in. The two formats have been presented as particularly suitable for high-performance streaming. However, Apache Jena only implements the most rudimentary, byte-level streaming support with Java IO Streams. This, obviously, is not appropriate for end-to-end RDF streaming.

\section{Description of the Proposed Approach}

Bearing in mind the outlined research goals, a solution was designed, consisting of a new RDF serialization format and integrations with two streaming protocols. The method makes extensive use of already available protocols, libraries, and standards. This has the major benefit of simplifying the presented reference implementation, as well as any future implementations for other software and hardware platforms. Moreover, it helps to avoid numerous performance and security pitfalls, associated with, e.g., implementing a new network protocol.

\subsection{Serialization Format}

The new \textsc{Jelly}\footnote{The name does not bear any meaning, beyond being a short for \textit{Superfast~Jellyfish}.} serialization format is based on the popular, high-performance Protocol Buffers format. It uses several strategies to reduce payload size, while maintaining very high performance. It supports serializing generalized RDF triples and quads, which allows it to cover all aspects of the RDF 1.1 specification~\cite{Cyganiak:14:RCA}. Moreover, the upcoming RDF-star~\cite{rdf-star} standard for reification is also supported.

\begin{figure}[tb]
\centerline{\includegraphics[width=8cm]{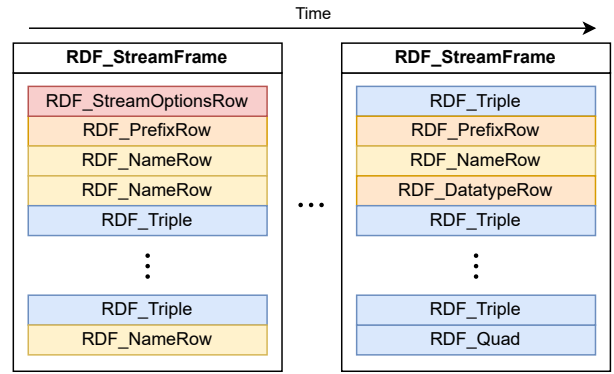}}
\caption{An example \textsc{Jelly} stream.}
\label{fig:frames}
\end{figure}

Let us now describe the details of the proposed serialization. The RDF stream is represented as a series of \texttt{RDF\_StreamFrame} objects, each containing a list of \texttt{RDF\_StreamRow} objects (Fig.~\ref{fig:frames}). The number of rows per frame is directly not limited by the proposed approach, and can be as low as one, and as high as the network protocol will allow. The size of the frame can also vary over the duration of the stream. The serialization depends on the streaming protocol to provide reliable transmission of messages, which is a feature now commonly found in many solutions, such as gRPC, Kafka, and MQTT. The frames and rows must be processed strictly in the order, in which they arrive. This is because preceding rows can set the context for the future rows. Both the serializer and deserializer must maintain a lightweight internal mutable state (containing, among others, three lookup tables), to be able to update this context dynamically. For streaming brokers (Kafka, MQTT) this places a requirement on the supported Quality of Service, delivery guarantees and message offsets.

An \texttt{RDF\_StreamRow} is an atomic stream element that can be one of:
\begin{itemize}
    \item \texttt{RDF\_StreamOptionsRow} -- a small header at the beginning of the stream that informs the consumer about the used compression settings.
    \item \texttt{RDF\_PrefixRow} -- a new entry in the prefix lookup table (explained below). New entries to lookup tables can be placed anywhere in the stream.
    \item \texttt{RDF\_NameRow} -- a new entry in the name lookup table.
    \item \texttt{RDF\_DatatypeRow} -- a new entry in the datatype lookup table.
    \item \texttt{RDF\_Triple} -- a triple consisting of three \texttt{RDF\_Term} objects.
    \item \texttt{RDF\_Quad} -- a quad, constructed analogically to a triple.
\end{itemize}

An \texttt{RDF\_Term} object can be one of:
\begin{itemize}
    \item \texttt{RDF\_IRI} -- an RDF IRI, expressed as a pair of identifiers, one for the prefix, and one for the name (remainder of the IRI). These refer to entries in the prefix and name lookup tables. Both identifiers are optional.
    \item \texttt{RDF\_BNode} -- a blank node with a unique string identifier.
    \item \texttt{RDF\_Literal} -- a literal with its lexical value (as a string) and, optionally, a language tag or a datatype tag. The datatype is an identifier referring to an entry in the datatype lookup table.
    \item \texttt{RDF\_Triple} –- a quoted triple in RDF-star.
    \item \texttt{RDF\_REPEAT} -- a special tag instructing the deserializer to repeat the term that was previously seen in this position (subject, object, predicate, or graph name) of the triple or quad.
\end{itemize}

One notable feature of the proposed format is the use of three, dynamically updated, lookup tables. They are integer-indexed and their maximum size is communicated by the producer to the consumer at the start of the stream (in the stream options row). The consumer can use this information to pre-allocate the necessary lookup tables, but it can also expand them dynamically as needed. The lookup table size, obviously, influences both the compression performance and the memory usage. Thus, it can be freely adjusted from as low as single-digit numbers to $2^{28}$, however, using such huge lookup tables would be impractical due to excessive memory consumption. The lookup table size is usually between 30 and 10000.

\begin{figure}[tb]
\centerline{\includegraphics[width=8cm]{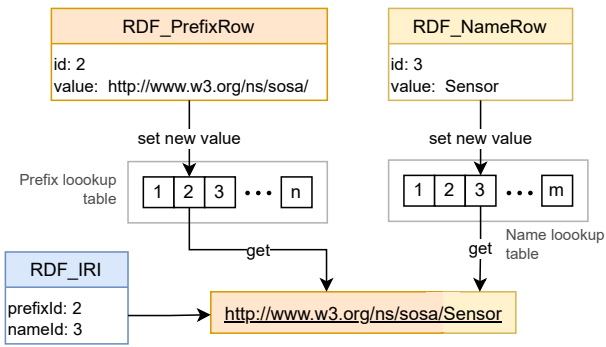}}
\caption{IRI reconstruction during deserialization.}
\label{fig:des_iri}
\end{figure}

The serializer is responsible for adding new elements to the lookup, and evicting old ones. Here, any algorithm can be applied to achieve this. In the reference implementation, a linked hash map with a least-recently-used (LRU) policy is used. It is a relatively uncomplicated algorithm that achieves reasonable compression efficiency, by dynamically replacing least recently used items in the lookup with new ones, when the need arises. These updates are communicated to the consumer with the \texttt{RDF\_PrefixRow}, \texttt{RDF\_NameRow}, and \texttt{RDF\_DatatypeRow} objects. The task of the deserializer is more straightforward. Its lookups can be implemented as plain arrays, with $\mathcal{O}(1)$ update and retrieve complexity (see Fig.~\ref{fig:des_iri}).

The \texttt{RDF\_REPEAT} term offers an efficient, $\mathcal{O}(1)$ compression mechanism, similar in its premise to the Turtle RDF serialization. By eliminating repeating terms, it not only reduces the size of the data to be transmitted, but also speeds up processing. This speedup is due to the serializer and deserializer not having to encode/decode the same term multiple times.

Overall, the available compression settings should allow the serialization to flexibly fit the needs of both high-performance servers and resource-constrained IoT devices. The lookup tables can be almost arbitrarily sized, and updated with any algorithm. For example, in an IoT device that produces only a stream of triples with fixed subjects and predicates, static (or partially static) lookup tables can be employed. The prefix table can be disabled entirely to simplify encoding and reduce memory requirements, at the cost of increased payload size. Conversely, in high-performance devices, larger lookup tables may be used, with more sophisticated eviction strategies (for example, least-frequently used). Additional compression (e.g., gzip) of the serialized messages can be introduced, to further reduce the size of the payload. This is, however, not within the scope of the serialization format itself, and is typically handled by the streaming protocol.

The serialization format is formalized using the Protocol Buffer language version 3. This specification can be used to quickly generate the needed serialization/deserialization boilerplate code, in any of the many supported languages. Currently, a reference implementation in Scala is provided~\cite{materials}. The implementation is integrated with the popular Apache Jena RDF library.

\subsection{Integration with Streaming Protocols}

The serialization format by itself is of little use -- it is necessary to integrate it with network streaming protocols. To fulfill this requirement, \textsc{Jelly} is integrated with two protocols that can be used for streaming purposes. The first is gRPC\footnote{\url{https://grpc.io/}} -- a Remote Procedure Call (RPC) framework, supporting bi-directional streaming, oriented towards synchronous communication between services. gRPC's services are defined using the Protocol Buffers Interface Definition Language, which allows for implementing code compliant with the protocol specification quickly, on many platforms. gRPC is based on the modern HTTP/2 standard, which natively supports gzip compression, with content type negotiation. The second integration is with Apache Kafka\footnote{\url{https://kafka.apache.org/}}, a popular asynchronous, distributed streaming platform. It possesses a number of advanced features that allow building highly scalable and resilient streaming applications. Both integrations are based on the Akka framework\footnote{\url{https://akka.io/}}.

It is important to note that although both gRPC and Kafka support streaming communication, they have very different architectures and are designed for different applications. In Section~\ref{eval}, one of the key aspects of the evaluation is addressing this very issue, to establish the strengths and weaknesses of each approach, when applied to RDF streaming.

One reason for this particular choice of streaming protocols is precisely due to them being very different. This helps demonstrate the method's applicability in a variety of scenarios. Secondly, both gRPC and Kafka are high-performance, mature solutions, that have seen wide adoption in the industry.

\section{Experimental Evaluation}\label{eval}

Selected aspects of the the proposed method have been evaluated in a comprehensive set of benchmarks. The tests were performed on a 14 vCore virtual machine (Intel Xeon E5-2640 v2) with 16\,GB of RAM, running Ubuntu 20.04.4 LTS, Linux kernel version 5.4.0. The Java Virtual Machine (JVM) used was GraalVM Community 22.1.0, running in the standard JIT mode.

The RDF streaming methods presented in Section~\ref{sec:bg} are not compared in most of the experiments reported here. The reasons are that (1) their implementations are not publicly available (\mbox{S-HDT}, EXI, CQUELS-Cloud, RDSZ), or (2) they are not integrated with any RDF library or network protocol (ERI). To make reliable streaming throughput and latency assessment, the compared methods should be based on the same RDF library. This is necessary to properly compare the performance of the full pipeline -- transforming in-memory objects to the serialized representation, and back. Similarly, access to the implementation is needed to compare the methods on the same test platform (hardware, operating system, etc.). As no such methods are available, \textsc{Jelly} is compared to the serialization formats already available in Apache Jena. For the purpose of the evaluation, Kafka stream connectors were provided for the built-in Jena serialization formats.

\subsection{Datasets Used in Experiments}

For the experiments, a subset of the datasets used in the evaluation of ERI~\cite{fernandez2014efficient} was selected, focusing on the streaming datasets. Not all were included, either because of their invalidity\footnote{For example, the LinkedMDB dataset contains invalid IRIs and cannot be loaded into Apache Jena.} or processing time limitations. Moreover, to avoid disk overhead influencing the benchmarks, all selected datasets had to fit into system memory. Thus, the Flickr\_Event\_Media and LOD\_Nevada datasets were trimmed to the the first 10 million triples. The following ten datasets were used: Mix, Identica, Wikipedia, \mbox{AEMET-1}, \mbox{AEMET-2}, Petrol, Flickr\_Event\_Media (10M), LOD\_Nevada (10M), Eurostat\_migr\_reschange, Eurostat\_tour\_cap\_nuts3. It is worth pointing out that these datasets were designed to be representative of a wide variety of semantic data streaming use cases. Thus, they are based on data gathered from real applications and services.

Unless otherwise stated, the following results were averaged over all ten datasets. Detailed per-dataset results can be found in the supplementary materials.

\subsection{Serialization and Deserialization Performance}

First, the raw serialization and deserialization performance was measured for \textsc{Jelly} (four variants of compression settings) and four other serializations included in Apache Jena. Jena's serializers can only work with RDF graphs, thus the stream of triples was transformed into a stream of 1000-triple graphs (even though RDF graphs are sets of triples, in the JVM environment not every Set whose elements are triples is a Jena Graph object). The benchmarks were ran on a single thread, with JVM being able to use additional threads, e.g., for just-in-time compilation and garbage collection. The following methods were tested:
\begin{itemize}
    \item \textbf{jelly-full} – \textsc{Jelly} with the default settings – prefix table size: 150, name table size: 4000, datatype table size: 32.
    \item \textbf{jelly-noprefix} – jelly-full, with the prefix table disabled.
    \item \textbf{jelly-noprefix-sm} – jelly-noprefix, with the name table size reduced to 256.
    \item \textbf{jelly-norepeat} – jelly-full, without the use of the \texttt{RDF\_REPEAT} term.
    \item \textbf{jena-protobuf} – Jena's Protocol Buffers serialization.
    \item \textbf{jena-n3} – W3C N-Triples format.
    \item \textbf{jena-turtle} – W3C Turtle format.
    \item \textbf{jena-xml} – W3C RDF/XML format.
\end{itemize}

The results can be found in Table~\ref{tab:raw_ser_des}. All values are indicated in thousands of triples processed per second (kT/s; highest values in bold). Overall, \textsc{Jelly} offers significantly faster deserialization than any Jena-based method, including its Protocol Buffers implementation. For serialization, the proposed method is visibly slower than Jena's Protocol Buffers implementation and much slower than the \mbox{N-Triples} format. This difference can be explained by the binary encoding and additional compression performed by \textsc{Jelly}. In serialization, the \textit{noprefix} and \textit{noprefix-sm} variants are the fastest, due to the simpler compression algorithm (no prefix lookup table). Among all variants, \textit{norepeat} is visibly the slowest, which is due to the additional burden of encoding the same term multiple times, instead of using an \texttt{RDF\_REPEAT} term.

In Table~\ref{tab:raw_ser_des}, the last column indicates the total theoretical throughput of the method, calculated as the smaller value of the serialization and deserialization speeds. This simple calculation obviously cannot be used as an indicator of how the methods would perform in real-life conditions. It is, however, interesting to note that, owing to their superior deserialization speeds, all variants of \textsc{Jelly} should, theoretically, outperform Jena's serializers. The text-based serializations (N3, Turtle, RDF/XML) are not competitive with the binary representations, with the notable exception of N3's serialization speed.

\begin{table}[tb]
\caption{Raw Ser/Des Performance (kT/s)}
\begin{center}
\begin{tabular}{|c|r|r|r|}
\hline
\textbf{Method} & \multicolumn{1}{|c|}{\textbf{Serialization}} & \multicolumn{1}{|c|}{\textbf{Deserialization}} & \multicolumn{1}{|c|}{\textbf{Th. throughput}} \\
\hline
jelly-full & 562.45 & 820.35 & 562.45 \\
jelly-noprefix & 752.92 & \textbf{889.99} & 752.92 \\
jelly-noprefix-sm & 765.66 & 882.75 & \textbf{765.66} \\
jelly-norepeat & 453.93 & 702.61 & 453.93 \\
\hline
jena-protobuf & 1060.07 & 465.04 & 465.04 \\
jena-n3 & \textbf{2664.29} & 148.05 & 148.05 \\
jena-turtle & 155.90 & 158.19 & 155.90 \\
jena-xml & 71.95 & 79.23 & 71.95 \\
\hline
\end{tabular}
\label{tab:raw_ser_des}
\end{center}
\end{table}

In further, end-to-end streaming experiments, the performance of these methods is examined in more realistic scenarios. However, the largely theoretical results presented here help establish the baseline performance for the investigated methods.

\subsection{Serialized Representation Size}

The size of the serialized representations was compared for \textsc{Jelly}, the Jena-based serializations, and two earlier methods (ERI~\cite{fernandez2014efficient} and the non-streaming HDT~\cite{fernandez2013binary}), based on the figures reported in the paper about ERI~\cite{fernandez2014efficient}. For Jena serializations, 1000-triple message size was assumed, the same as in the previous experiment. For ERI and HDT only the figures for 4096-triple messages were available. Additionally, each method was tested with a supplementary gzip compression stage.

The results can be found in Table~\ref{tab:size}. The compression ratios are calculated as the ratio between the serialized size of the representation and the size for the equivalent uncompressed \mbox{N-Triples} file. The reported figures are the geometric means of the ratios on all ten datasets. It can be observed that \textsc{Jelly}, when used without gzip, achieves much better results than RDF/XML or Turtle, while being vastly faster than either of them. With gzip, its compression ratios are comparable to the Jena-based results, and noticeably worse than ERI, but better than HDT. The differences between the different variants of the proposed method are relatively small. This suggests that, for example, using the prefix table is useful only in constrained network environments, or when working with data that lends itself well to this particular compression method.

The Jena Protocol Buffers-based representation achieves very poor compression ratios, with larger payloads than \mbox{N-Triples} (when used without gzip). ERI offers the best compression among all tested approaches. However, as it was not integrated with an RDF library, it could not be examined in the other, performance-oriented experiments. Moreover, the larger message size (4096 vs 1000) likely gives an advantage to both ERI and HDT. Consequently, it is impossible to say if ERI and HDT would outperform other approaches in practice. It is crucial to remember that the representation size is just one of the many variables influencing streaming performance and, as evidenced in the following sections, it is by itself a poor indicator of how the method will perform.
\begin{table}[tb]
  \centering
  \caption{Serialized Size Comparison for Jelly and Competing Methods.}
  \begin{tabular}{|c|c|r|}
    \hline
    \multicolumn{2}{|c|}{\textbf{Method}} & \multirow{2}{*}{\shortstack{\textbf{Compression ratio}\\ \textbf{(geom. mean)}}} \\
    \cline{1-2}
    \textbf{Serialization} & \textbf{gzip} & \\
    \hline
    jelly-full & $-$ & 25.55\,\% \\
    jelly-full & $+$ & 3.41\,\% \\
    jelly-noprefix & $-$ & 27.57\,\% \\
    jelly-noprefix & $+$ & 3.44\,\% \\
    jelly-noprefix-sm & $-$ & 29.48\,\% \\
    jelly-noprefix-sm & $+$ & 3.69\,\% \\
    jelly-norepeat & $-$ & 29.10\,\% \\
    jelly-norepeat & $+$ & 5.49\,\% \\
    \hline
    jena-protobuf & $-$ & 108.33\,\% \\
    jena-protobuf & $+$ & 6.71\,\% \\
    jena-n3 & $-$ & 100.00\,\% \\
    jena-n3 & $+$ & 5.80\,\% \\
    jena-turtle & $-$ & 72.20\,\% \\
    jena-turtle & $+$ & 3.20\,\% \\
    jena-xml & $-$ & 50.05\,\% \\
    jena-xml & $+$ & 2.99\,\% \\
    \hline
    ERI-4k & N/A & *\,2.23\,\% \\
    ERI-4k-Nodict & N/A & *\,2.11\,\% \\
    HDT-4k & N/A & *\,6.71\,\% \\
    \hline
    \multicolumn{3}{p{0.63\linewidth}}{*\,Estimated from the results reported in the paper on ERI~\cite{fernandez2014efficient}.}
  \end{tabular}
  \label{tab:size}
\end{table}

\subsection{End-to-End Streaming Throughput}

To evaluate the practical achievable performance of the method, it was tested in tandem with two streaming protocols (gRPC and Kafka), over a full network stack. Additionally, two scenarios of limited networking were simulated, using the NetEm Linux kernel module~\cite{hemminger2005network}, by introducing a limit on the bandwidth and an artificial latency for the packets. The first scenario had 10\,ms latency and 100\,Mbit/s bandwidth (10\,/\,100) and the other had 15\,ms latency and 50\,Mbit/s bandwidth (15\,/\,50). To maintain a possibly fair comparison, in tests involving Kafka, only the connection between the producer and the broker was throttled -- the consumer had always unlimited networking. For Kafka, to avoid issues with rebalancing and ``warmup'' of the broker, a single cluster, with one consumer group and single partition was used. Other settings were left at their defaults, with at-most-once delivery guarantees from Kafka.

The results are presented in Table~\ref{tab:full_throughput}. In all network conditions \textsc{Jelly} vastly outperforms Jena-based serializations, which is in line with the results from the previous experiment on theoretical throughput. It can be observed that gzip slows down streaming in unlimited network conditions, but as soon as bandwidth limitations are introduced, it becomes indispensable for Kafka streaming. For gRPC this effect is only visible with 50\,Mbit/s networking. Finally, Kafka can generally provide better overall throughput than gRPC. However, this advantage seems to shrink, when network bandwidth becomes limited.
\begin{table}[htbp]
\caption{End-to-End Streaming Throughput (kT/s)}
\begin{center}
\begin{tabular}{|c|c|c|r|r|r|}
\hline
\multicolumn{3}{|c|}{\textbf{Method}} & \multicolumn{3}{|c|}{\textbf{Network conditions}} \\
\hline
& \textbf{Serialization} & \textbf{gzip} & \multicolumn{1}{|c|}{\textbf{Unlimited}} & \multicolumn{1}{|c|}{\textbf{10\,/\,100}} & \multicolumn{1}{|c|}{\textbf{15\,/\,50}} \\
\hline
\parbox[t]{2mm}{\multirow{4}{*}{\rotatebox[origin=c]{90}{\textbf{gRPC}}}} & jelly-full & $-$ & 398.55 & 292.91 & 150.12 \\
& jelly-full & $+$ & 270.11 & 274.45 & 268.22 \\
& jelly-noprefix & $-$ & 482.51 & 269.76 & 138.43 \\
& jelly-noprefix & $+$ & 314.91 & 311.45 & \textbf{311.79} \\
\hline
\parbox[t]{2mm}{\multirow{8}{*}{\rotatebox[origin=c]{90}{\textbf{Kafka}}}} & jelly-full & $-$ & 480.73 & 56.67 & 35.91 \\
& jelly-full & $+$ & 322.18 & 292.69 & 272.03 \\
& jelly-noprefix & $-$ & \textbf{597.55} & 52.75 & 33.34 \\
& jelly-noprefix & $+$ & 391.37 & \textbf{342.58} & 304.86 \\
\cline{2-6}
& jena-protobuf & $-$ & 282.33 & 18.74 & 11.85 \\
& jena-protobuf & $+$ & 257.09 & 227.28 & 186.61 \\
& jena-n3 & $-$ & 125.15 & 20.16 & 12.73 \\
& jena-n3 & $+$ & 126.49 & 120.35 & 123.27 \\
\hline
\end{tabular}
\label{tab:full_throughput}
\end{center}
\end{table}

\subsection{End-to-End Streaming Latency}

Maintaining low latency is crucial in many streaming applications, including IoT, often making the difference between the technology being feasible or not~\cite{schulz2017latency}. Thus, the streaming latency of the proposed method was examined in a variety of settings. In each test, a stream was opened between the producer and the consumer. Next, 1000 messages were published at regular intervals. The latency for each message was measured as the time between it entering the serializer and it leaving the deserializer (end-to-end). All combinations of the following settings were tested:
\begin{itemize}
    \item Network: three different setups, same as in the previous experiment.
    \item Message size: 5, 50, 200 triples.
    \item Interval between messages: 100\,\textmu s, 1\,ms, 10\,ms.
    \item Dataset: Wikipedia, AEMET-1, AEMET-2, Petrol, Flickr\_Event\_Media (10M)
\end{itemize}

\begin{figure}[tb]
    \centering
    \includegraphics[width=\linewidth]{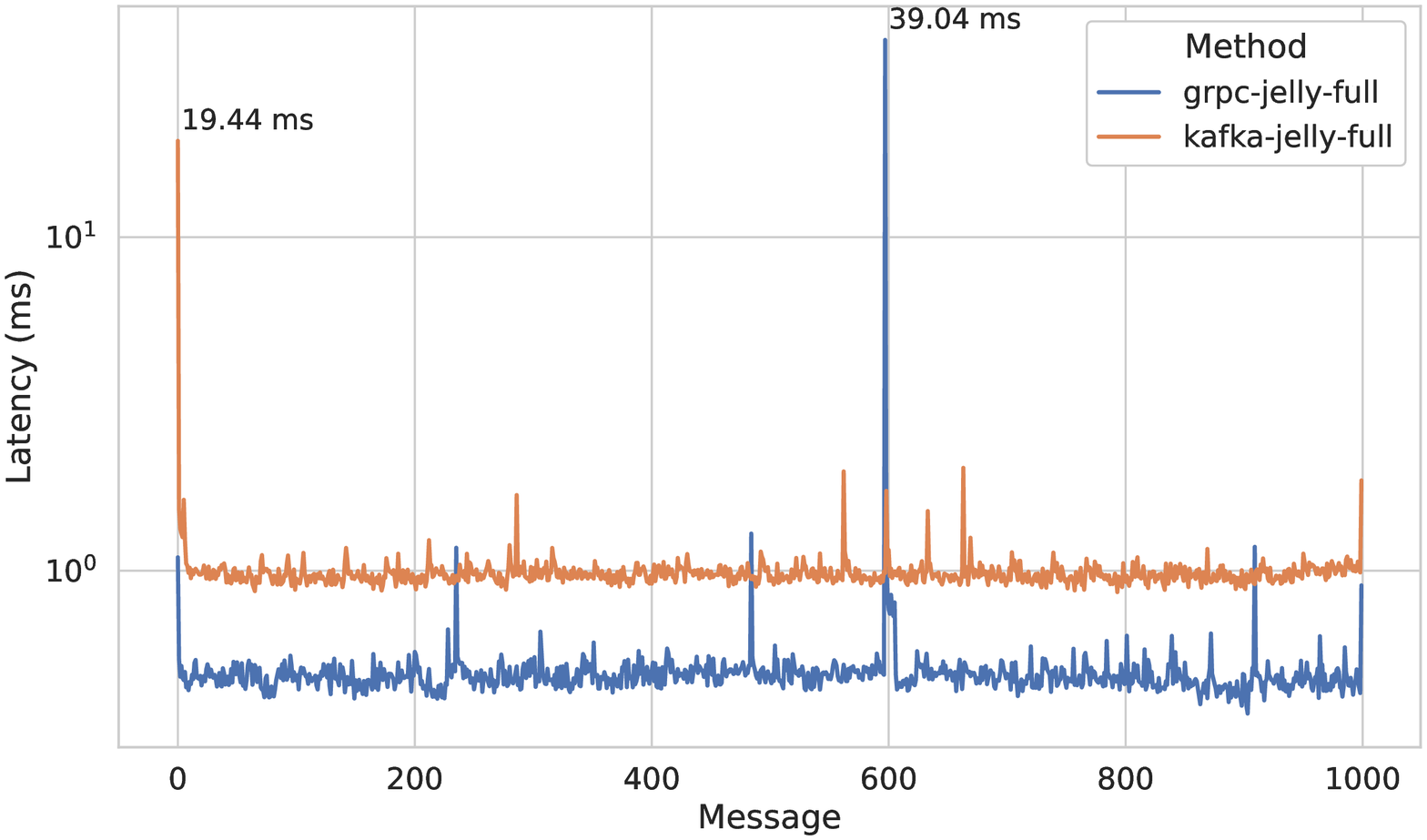}
    \caption{Streaming latency of 1000 consecutive messages (network:~unlimited, interval:~100\,\textmu s, message size: 50, dataset: \mbox{AEMET-1})}
    \label{fig:latency_series}
\end{figure}

Each run produced a series of 1000 values, an example of which is presented in Fig.~\ref{fig:latency_series}. Not all results could be presented here, due to their excessive volume, however, they are available in the supplementary materials~\cite{materials}.

\begin{figure*}[p]
    \centering
    \includegraphics[width=\textwidth]{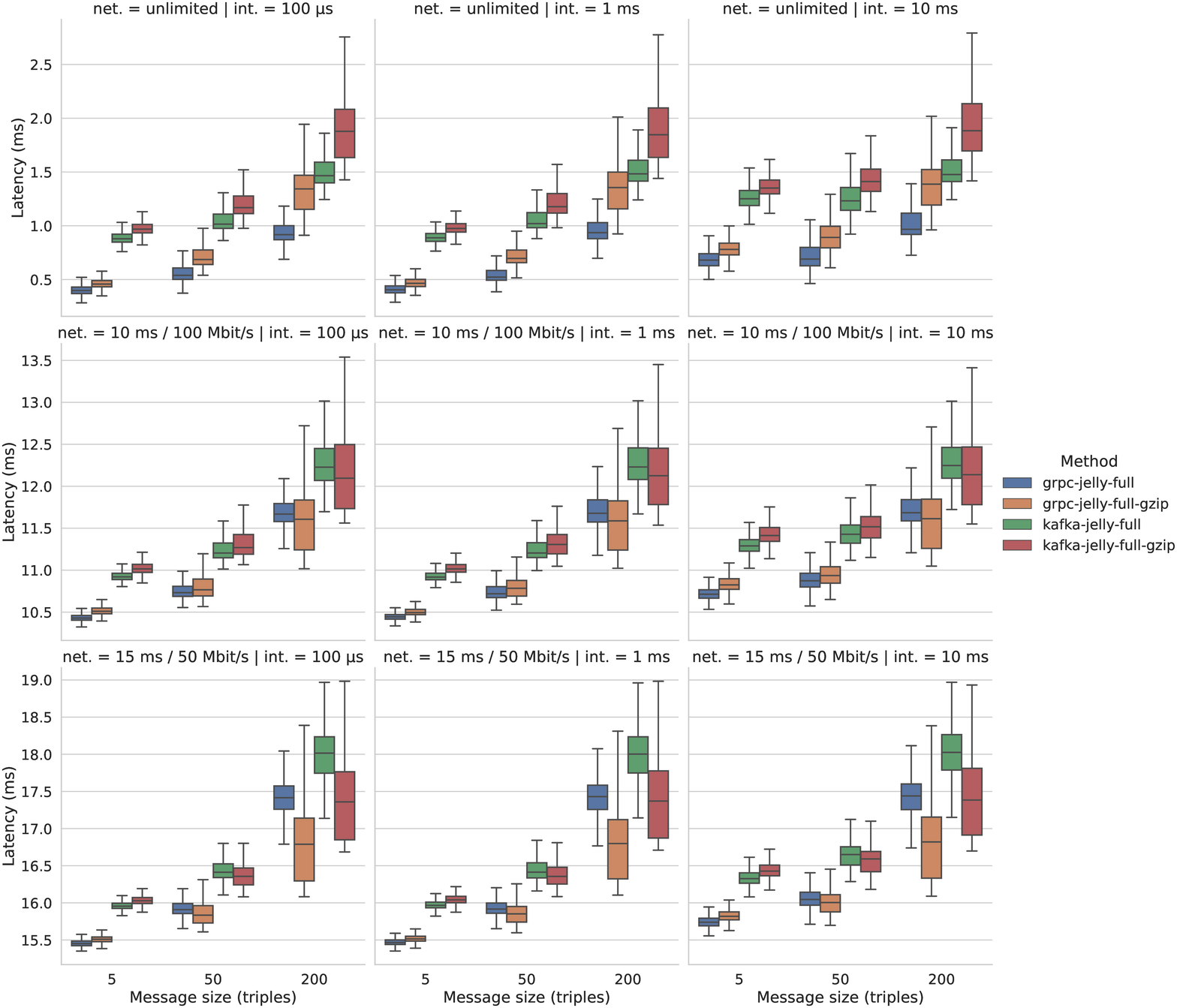}
    \caption{Influence of gzip compression and network conditions on streaming latency.}
    \label{fig:latency_gzip}
\end{figure*}

Fig.~\ref{fig:latency_unl} compares the performance of non-gzip serializations in unlimited network conditions. It can be clearly seen that \textsc{Jelly} provides much lower latencies than Jena-based serializations, and that gRPC outperforms Kafka in all of the presented cases. The same observations hold true for limited networking conditions (not presented here). Fig.~\ref{fig:latency_gzip} presents how gzip compression impacts latencies in limited and unlimited networks. Here, it is apparent that gzip increases latency in unconstrained network conditions, which is expected, as it necessitates additional processing. However, in a 50\,Mbit/s network, gzip starts giving a clear advantage for messages that are large enough to benefit from the additional compression.

Overall, it can be observed that the combination of \textsc{Jelly} and gRPC results in sub-one millisecond latencies, greatly improving over Kafka, and the standard serializations.

\begin{figure*}[p]
    \centering
    \includegraphics[width=\textwidth]{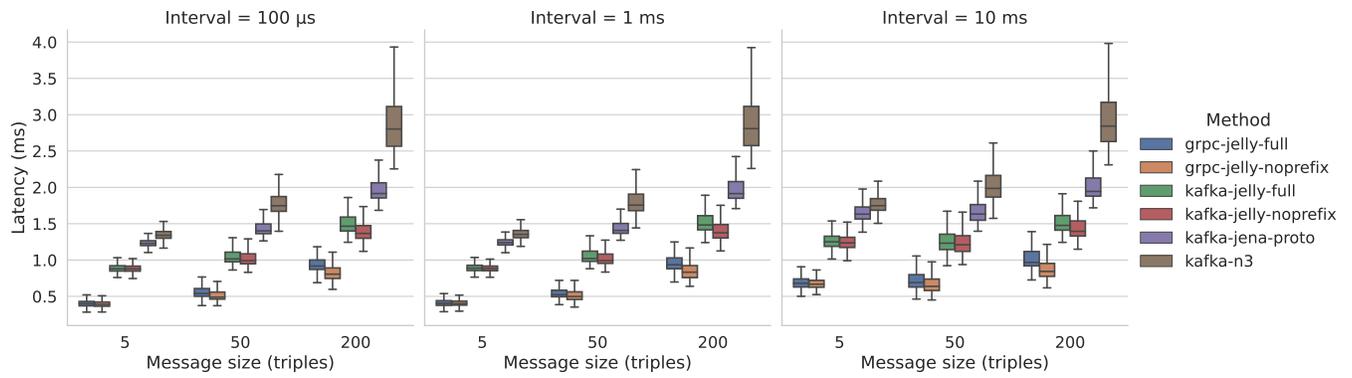}
    \caption{Streaming latency in unlimited network conditions.}
    \label{fig:latency_unl}
\end{figure*}

\subsection{Evaluation Limitations}

It is important to note the limitations of the above evaluation. Firstly, in end-to-end streaming both the producer and the consumer were running on the same virtual machine and the same JVM, which may result in resource contention. On the other hand, this setup has allowed for sub-microsecond precision in measuring throughput and latency. Secondly, the results could be influenced by any background tasks that have been scheduled by the JVM, such as garbage collection and JIT compilation. To mitigate this, the tests include a few ``warm-up'' runs and are then repeated several times to gather more representative data.

\section{Conclusions and Future Work}

The presented approach makes significant progress towards enabling practical RDF data streaming in IoT and future edge-cloud systems, by meeting the demands for high performance (as evidenced in the exhaustive performance evaluation), scalability (by being configurable), and portability (by being simple to implement, and by the use of established technologies and standards). \textsc{Jelly} advances the current state of the art, being the first end-to-end, publicly available RDF streaming solution. This is in stark contrast to the current state-of-the-art solutions, which solve only parts of the problem, and are (most often) closed-source, which hampers the development of future semantic data streaming systems. The proposed method provides high performance and scalability in a variety of scenarios, while being relatively simple to implement. It was integrated with two modern communication protocols, and a popular RDF framework, making it readily applicable.

There remains much to be explored in the topic of semantic data streaming. In the future, the performance of \textsc{Jelly} will be evaluated in a variety of real world-anchored use cases, including those pertaining to edge-cloud systems. Namely, it is planned to be used in the currently ongoing ASSIST-IoT EU Horizon 2020 project, to meet the demands for low-latency and high-throughput RDF streaming. Furthermore, it is planned to improve the implementation of the protocol and create new implementations to facilitate \textsc{Jelly}'s reuse on various platforms. Currently considered is the popular Python rdflib library, and a C++ implementation for microcontrollers. As for streaming protocols, an integration with the MQTT standard\footnote{\url{https://mqtt.org/}} is planned, due to its ubiquity in modern IoT ecosystems.

\vfill\null

\bibliographystyle{IEEEtran}
\bibliography{bibliography}

\end{document}